\documentclass[reprint,twocolumn,prl,showpacs,showkeys,aps]{revtex4-1}
\usepackage{epsfig,amsmath,amsfonts,amssymb}
\usepackage{amsmath,amsfonts,amssymb}
\usepackage{graphicx}
\usepackage{subfigure}
\usepackage{mathbbol}
\usepackage{amsthm}

\begin{document}

\title{Efficient entanglement generation between exciton-polaritons using shortcuts to adiabaticity}

\author{Dionisis Stefanatos}
\email{dionisis@post.harvard.edu}
\author{Emmanuel Paspalakis}%
%\altaffiliation[Current address:]{ 3 Omirou St., Sami, Kefalonia 28080, Greece.}
%\email{dionisis@post.harvard.edu}
\affiliation{Materials Science Department, School of Natural Sciences, University of Patras, Patras 26504, Greece}

\date{\today}% It is always \today, today,

\begin{abstract}
We use shortcuts to adiabaticity, a method introduced to speed up adiabatic quantum dynamics, for the efficient generation of entanglement between exciton-polaritons in coupled semiconductor microcavities. A substantial improvement is achieved, compared to a recently proposed method which essentially enhances the nonlinearity of the system. Our method takes advantage of a time-dependent nonlinearity which can become larger than the Josephson coupling between the cavities, while the conventional method is restricted to a constant nonlinearity lower than the coupling. The suggested procedure is expected to find also application in other research areas in optics, where nonlinear interacting bosons are encountered.
\end{abstract}

%\pacs{}
%\pacs{05.45.Xt}% PACS, the Physics and Astronomy
%\keywords{quantum control}%Use showkeys class option if keyword
\maketitle

\textbf{Introduction}: Exciton-polaritons are bosonic quasiparticles in semiconductor microcavities formed by the strong coupling between excitons and cavity photons \cite{Byrnes14}, which can be easily manipulated and loaded to two-dimensional trap arrays forming lattices. These structures of nonlinear interacting quantum oscillators provide an attractive solid state platform for quantum information processing with continuous variables. Current applications include the generation of non-classical states of light \cite{Liew10,Klaas18}, qubits and gates for quantum computation \cite{Demirchyan14,Kyriienko16}, and quantum simulators \cite{Askitopoulos15,Ohadi17,Sigurdsson17,Berloff17,Lagoudakis17}.

One extremely important task regarding the use of these systems in quantum information, is the efficient generation of entanglement between exciton-polaritons in coupled cavities. The problem is that entanglement creation relies on the strength of the nonlinearity, which is weak for semiconductor microcavities. Entanglement appears only as a perturbation \cite{Sun17,Stefanatos18} even at high densities, where the nonlinear effects become important, since the mean field approximation provides a fair classical description  \cite{Casteels17}. In order to overcome this barrier, a method to essentially amplify the nonlinearity strength in semiconductor microcavities using two coherent laser fields was recently suggested \cite{Liew18}, leading in theory to the creation of a fair amount of entanglement between exciton-polaritons in coupled cavities. Throughout this process, the strength of enhanced nonlinearity in the two coupled cavities is held constant and smaller than the Josephson coupling.

In this work, we consider a time-dependent enhanced nonlinearity, through the modulation of the corresponding coherent laser fields, which is also allowed to attain larger values than the Josephson coupling. Under this setting, we use shortcuts to adiabaticity (STA), a method developed to accelerate quantum adiabatic dynamics, to efficiently generate entanglement between two coupled cavities. The essence of STA is that it drives the system to the same final state as a slow adiabatic process but in a much shorter time, since it does not necessarily follow the instantaneous eigenstates. It has been exploited to efficiently perform various tasks in several fields of optics \cite{Garaot14,Tseng14,Stefanatos14,Ho15,Guo17}. Here, we show that STA can take advantage of the large enhanced nonlinearity and lead to a substantially larger amount of entanglement, compared to that obtained in the recent work \cite{Liew18}. %The presented methodology is not restricted only to exciton-polaritons in semiconductor microcavities, but is expected to have a broader impact in optics research since the model of two nonlinear interacting bosons is encountered in a wide spectrum of physical settings.

\textbf{Model}: We consider a pair of coupled cavities as in \cite{Liew18}, which can be implemented with the techniques of \cite{Vasconcelos11}, described by the Hamiltonian
\begin{equation}
\label{Hamiltonian}
\hat{H}=\frac{\alpha(t)}{2}(\hat{a}_1^2+\hat{a}_2^2+\hat{a}_1^{\dagger 2}+\hat{a}_2^{\dagger 2})-J(\hat{a}_1^{\dagger}\hat{a}_2+\hat{a}_1\hat{a}_2^{\dagger}).
\end{equation}
The first part of the Hamiltonian originates from an inverse four-wave mixing process in each cavity. As discussed in Ref. \cite{Liew18}, the starting point is the Hamiltonian $H_i=\alpha_0(\hat{a}_i^{\dagger}\hat{a}_i^{\dagger}\hat{a}_L\hat{a}_U+\hat{a}_L^\dagger\hat{a}_U^\dagger\hat{a}_i\hat{a}_i)/2$, where $\alpha_0$ is the strength of this nonlinear process, typically weak compared to the dissipation rate $\Gamma$ in optical systems, while modes $\hat{a}_L,\hat{a}_U$ are driven by coherent laser fields, which can be described classically. Pairs of particles scatter from $\hat{a}_L,\hat{a}_U$ to mode $\hat{a}_i$ and we are left with the first part of Hamiltonian (\ref{Hamiltonian}), where $\alpha=\alpha_0\langle a_L\rangle\langle a_U\rangle$ is the nonlinearity enhanced by the classical field amplitudes, which can reach the regime $\alpha\gg\Gamma$. This is the advantage over the usual method where the central mode $\hat{a}_i$ is excited and correlations are created between $\hat{a}_L,\hat{a}_U$ \cite{Schwendimann03,Karr04R,Portolan09}, which requires a nonlinearity $\alpha_0$ stronger than the dissipation rate. The authors of Ref. \cite{Liew18} studied the system dynamics with a constant enhanced nonlinearity $\alpha$. Here, we consider that the classical field amplitudes $\langle a_L\rangle, \langle a_U\rangle$ can be varied with time, allowing for a time-dependent $\alpha(t)$. The second part of the Hamiltonian is the familiar Josephson coupling, with a constant coupling coefficient $J$.

\textbf{Entanglement quantification}: For a system of two oscillators coupled with a quadratic Hamiltonian like (\ref{Hamiltonian}) and starting from vacuum, the states are Gaussian and characterized by the covariance matrix $V$ of the corresponding position and momentum operators $\hat{q}_i, \hat{p}_i, i=1,2$. If we define $(\hat{x}_1,\hat{x}_2,\hat{x}_3,\hat{x}_4)=(\hat{q}_1,\hat{p}_1,\hat{q}_2,\hat{p}_2)$, the elements of the covariance matrix become $V_{ij}=\langle\hat{x}_i\hat{x}_j+\hat{x}_j\hat{x}_i\rangle/2$, where note that the first moments are zero due to the vacuum initial conditions.
The state of the system is described, up to a possibly time-dependent phase factor, by the following Wigner quasiprobability distribution \cite{Adesso04}
\begin{equation}
\label{Wigner}
W(\mathbf{x})=\frac{1}{4\pi^2\sqrt{V}}e^{-\frac{1}{2}\mathbf{x}^TV^{-1}\mathbf{x}},
\end{equation}
where $\mathbf{x}=(q_1,p_1,q_2,p_2)$ is the vector of phase-space variables.

Using the relations $\hat{q}_i=(\hat{a}_i+\hat{a}_i^\dagger)/\sqrt{2}, \hat{p}_i=i(\hat{a}_i^\dagger-\hat{a}_i)/\sqrt{2}$, the elements of the covariance matrix can be expressed in terms of the second moments of the creation and annihilation operators of the two resonators, for example $\langle\hat{a}_1^\dag\hat{a}_1\rangle, \langle\hat{a}_2^\dag\hat{a}_2\rangle, \langle\hat{a}_1^\dag\hat{a}_2\rangle, \langle\hat{a}_1^2\rangle$ etc., while the first moments are zero due to the initial conditions. Instead of using directly the second moment operators, we can use specific linear combinations of them, a set of ten operators introduced by Dirac to describe exactly two coupled quantum oscillators \cite{Dirac63}, which are the generators of the symplectic group $Sp(4)$. Under the evolution described by Hamiltonian (\ref{Hamiltonian}), a closed set of differential equations can be obtained for the expectation values of these operators. Specifically, they are actually grouped into subsystems which are linear and homogenous in these variables. When starting from vacuum, only the subsystem formed by the following three operators
\begin{subequations}
\label{operators}
\begin{eqnarray}
\hat{S}_1&=&\frac{1}{2}(\hat{a}_1^\dagger\hat{a}_1+\hat{a}_2\hat{a}_2^\dagger),\label{S1}\\
\hat{S}_2&=&\frac{i}{4}(\hat{a}_1^{\dagger 2}+\hat{a}_2^{\dagger 2}-\hat{a}_1^2-\hat{a}_2^2),\label{S2}\\
\hat{S}_3&=&\frac{1}{2}(\hat{a}_1^{\dagger}\hat{a}_2^{\dagger}+\hat{a}_1\hat{a}_2).\label{S3}
\end{eqnarray}
\end{subequations}
has nonzero initial conditions; the rest of the operators remain zero throughout and can be ignored.
Using Ehrenfest theorem for operators without explicit time dependence $d\langle \hat{A}\rangle/dt=\imath[H,\hat{A}]$ ($\hbar=1$), we find that the corresponding expectation values $S_i=\langle\hat{S}_i\rangle$, $i=1,2,3$ satisfy the following system of equations
\begin{subequations}
\label{system}
\begin{eqnarray}
\dot{S}_1&=&-2\alpha S_2,\label{system1}\\
\dot{S}_2&=&-2\alpha S_1+2JS_3,\label{system2}\\
\dot{S}_3&=&-2JS_2,\label{system3}
\end{eqnarray}
\end{subequations}
with initial conditions
\begin{equation}
\label{Initial_Conditions}
S_1(0)=1/2,\quad S_2(0)=S_3(0)=0.
\end{equation}
Under the above evolution, the following constant of the motion can be easily verified
\begin{equation}
\label{constant}
S_1^2-S_2^2-S_3^2=1/4.
\end{equation}

The covariance matrix $V$ can be expressed in terms of the nonzero values $S_i, i=1,2,3$ as
\begin{equation*}
V=
\left(
\begin{array}{cc}
  A & C \\
  C^T & B
\end{array}
\right)=
\left(
\begin{array}{cc|cc}
  S_1 & S_2 & S_3 & 0 \\
  S_2 & S_1 & 0 & -S_3 \\
  \hline
  S_3 & 0 & S_1 & S_2 \\
  0 & -S_3 & S_2 & S_1
\end{array}
\right).
\end{equation*}
In order to quantify entanglement we will use the logarithmic negativity, a quantity which for two-mode Gaussian states (\ref{Wigner}) actually measures the squeezing of appropriate field quadratures \cite{Adesso04}. For this particular case the logarithmic negativity is given by $\mathcal{N}=\mbox{max}[0, -\ln(2\tilde{\nu}_-)]$, where $\tilde{\nu}_-$ is the smallest symplectic eigenvalue of a modified covariance matrix $\tilde{V}$ corresponding to the partially transposed state. We can evaluate $\tilde{\nu}_-$ in terms of $S_i$ using the formula \cite{Adesso04}
\begin{equation*}
%\label{symplectic}
\tilde{\nu}_-=\sqrt{\frac{\tilde{\Delta}(V)-\sqrt{\tilde{\Delta}^2(V)-4\mbox{det}V}}{2}},
\end{equation*}
where $\tilde{\Delta}(V)=\mbox{det}A+\mbox{det}B-2\mbox{det}C=2(S_1^2-S_2^2+S_3^2)$ and $\mbox{det}V=(S_1^2-S_2^2-S_3^2)^2$ ,
from which we obtain
\begin{equation*}
%\label{symplectic_value}
\tilde{\nu}_-=\sqrt{S_1^2-S_2^2}-|S_3|<1/2.
\end{equation*}
The last inequality can be proved using (\ref{constant}), and the logarithmic negativity is given by the expression
\begin{equation}
\label{negativity}
\mathcal{N}=-\ln(2\tilde{\nu}_-)=-\ln\left[2\left(\sqrt{S_1^2-S_2^2}-|S_3|\right)\right].
\end{equation}
Using the constant of the motion (\ref{constant}), the expression (\ref{negativity}) for the logarithmic negativity becomes
\begin{equation}
\label{S3_negativity}
\mathcal{N}=\ln\left[2\left(\sqrt{S_3^2+1/4}+|S_3|\right)\right],
\end{equation}
which is an increasing function of $|S_3|$.

\begin{figure}[t]
\centering
\fbox{\subfigure[\ No dissipation $\Gamma=0$]{
\label{fig:nodissipation}
\includegraphics[scale=0.28]{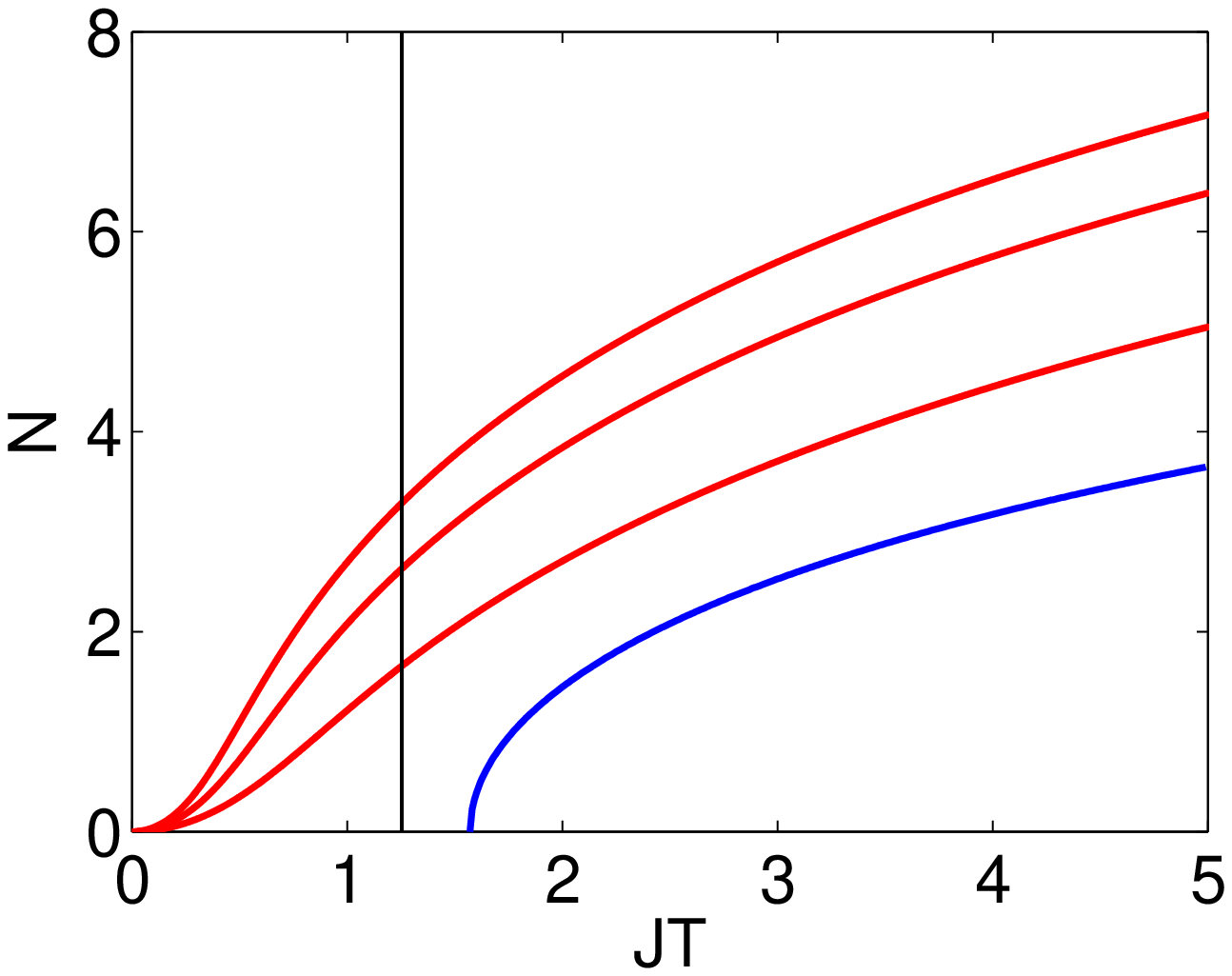}}
\subfigure[\ Dissipation $\Gamma=0.1J$]{
\label{fig:dissipation}
\includegraphics[scale=0.28]{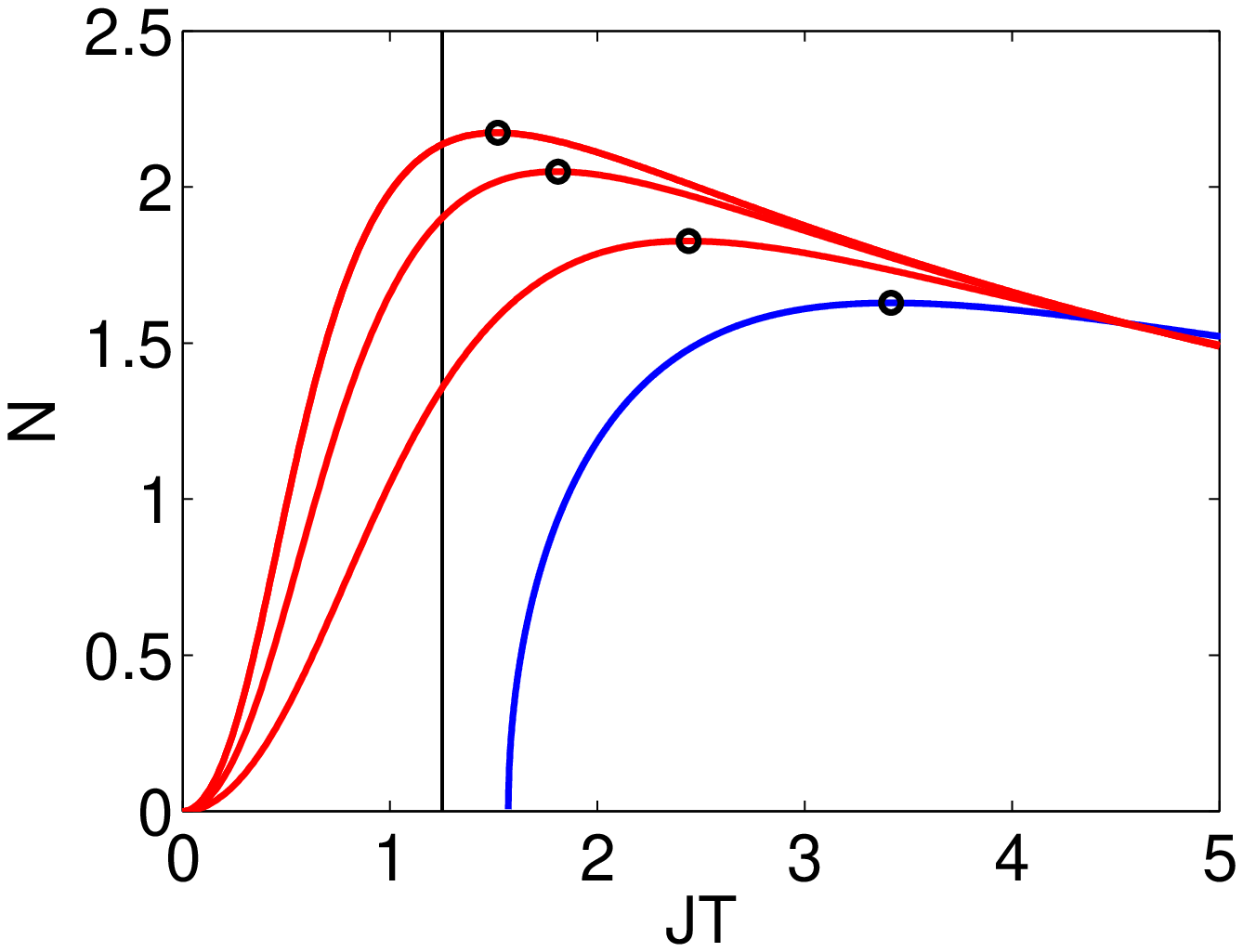}}}
\caption{Logarithmic negativity as a function of duration $JT$. (a) In the absence of dissipation, $\Gamma=0$. The lower blue curve corresponds to constant $\alpha(t)=\alpha_T$ given in (\ref{u_T}). The three upper red lines correspond to the maximum negativity obtained with the shortcut to adiabaticity control (\ref{alpha}) for three different upper bounds $\alpha(t)/J\leq A_0/J=2,4,6$ (from bottom to top) (b) Same as in (a) but in the presence of dissipation $\Gamma=0.1J.$}
\label{fig:negativity}
\end{figure}

\textbf{Constant enhanced nonlinearity}: In Ref.\ \cite{Liew18} a constant $\alpha(t)=\alpha_T$, with $\alpha_T<J$, is applied for the whole time interval $0\leq t\leq T$. The appropriate value of $\alpha_T$ depends on $T$, as it is denoted by the subscript.
By taking the time derivative of (\ref{system2}) and using (\ref{system1}), (\ref{system3}) we obtain the following differential equation for $S_2$
\begin{equation}
\label{Diff_S2}
\ddot{S}_2+4(J^2-\alpha_T^2)S_2=0.
\end{equation}
Solving for the initial conditions $S_2(0)=0, \dot{S}_2(0)=-\alpha_T$ we find
\begin{equation}
\label{solution_constant_a}
S_2(t)=-\frac{\alpha_T\sin(2\omega t)}{2\omega},\quad S_3(t)=\frac{J\alpha_T[1-\cos(2\omega t)]}{2\omega^2},
\end{equation}
where the angular frequency is $\omega=\sqrt{J^2-\alpha_T^2}$.
The choice
\begin{equation}
\label{u_T}
2\omega T=\pi\Rightarrow \alpha_T=\sqrt{J^2-\left(\frac{\pi}{2T}\right)^2}<J,
\end{equation}
leads to $S_2(T)=0, S_3(T)=J\alpha_T/\omega^2$. From Eq. (\ref{S3_negativity}) we find the logarithmic negativity as a function of the final time $T$
\begin{equation}
\label{old_negativity}
\mathcal{N}_T=2\ln\left[\frac{2JT}{\pi}+\sqrt{\left(\frac{2JT}{\pi}\right)^2-1}\right],
\end{equation}
and we plot it in Fig. \ref{fig:nodissipation} (lower blue curve). Observe that $T\geq\pi/(2J)$, while in the limit of large $T$ the logarithmic negativity increases logarithmically with time.

\textbf{Time-dependent enhanced nonlinearity}: We now find a smooth control $\alpha(t)$ which drives the system from the initial vacuum state to an eigenstate of the final Hamiltonian $\hat{H}(T)$, with a desired logarithmic negativity $\mathcal{N}$. Consider the following time-dependent operator
\begin{equation}
\label{I}
\hat{I}(t)=S_1(t)\hat{S}_1-S_2(t)\hat{S}_2-S_3(t)\hat{S}_3,
\end{equation}
where the operators $\hat{S}_i, i=1,2,3$ are given in Eqs. (\ref{S1})-(\ref{S1}), while their expectation values $S_i(t)$ satisfy system (\ref{system1})-(\ref{system3}).
It can be easily verified that $\hat{I}(t)$ satisfies
\begin{equation}
\label{invariant}
\frac{d\hat{I}}{dt}=\frac{\partial\hat{I}(t)}{\partial t}+i[\hat{H}(t),\hat{I}(t)]=0,
\end{equation}
thus it is a time-dependent invariant of motion. The initial vacuum state $|\Psi(0)\rangle=|00\rangle$ is an eigenstate of $\hat{I}$ at $t=0$, $\hat{I}(0)|\Psi(0)\rangle=(1/4)|\Psi(0)\rangle$. Let $|\phi_0(t)\rangle$ be the eigenstate of $\hat{I}(t)$ corresponding to the constant eigenvalue $\lambda_0=1/4$, i.e. $\hat{I}(t)|\phi_0(t)\rangle=\lambda_0|\phi_0(t)\rangle$. Then, by essentially taking the Fourier transform of the Wigner function (\ref{Wigner}), we find
\begin{equation}
\label{eigenstate}
\phi_0(\mathbf{q},t)=\frac{1}{\sqrt{2\pi}\sqrt[4]{S_1^2-S_3^2}}e^{iS_2\mathbf{q}^TK\mathbf{q}}e^{-\frac{1}{2}\mathbf{q}^TK\mathbf{q}},
\end{equation}
where $\mathbf{q}=(q_1, q_2)$ is the vector of coordinates and
\begin{equation}
\label{A}
K=\frac{1}{2(S_1^2-S_3^2)}
\left(
\begin{array}{cc}
  S_1 & -S_3 \\
  -S_3 & S_1
\end{array}
\right).
\end{equation}
It can be directly verified, by expressing the operators $\hat{S}_i, i=1,2,3$ in terms of the coordinates using the relations $\hat{a}_i=(q_i+\partial/\partial q_i)/\sqrt{2}, \hat{a}_i^\dagger=(q_i-\partial/\partial q_i)/\sqrt{2}, i=1,2$, that the above wavefunction is indeed eigenstate of $\hat{I}(t)$ with constant eigenvalue $\lambda_0=1/4$.
According to the theory of Lewis-Riesenfeld invariants \cite{Levy18}, the state of the system at time $t$ can be expressed as
\begin{equation}
\label{state}
|\Psi(t)\rangle=e^{i\theta_0(t)}|\phi_0(t)\rangle,
\end{equation}
where the phase $\theta_0$ is chosen as
\begin{equation}
\label{phase}
\frac{d\theta_0}{dt}=\langle\phi_0(t)|i\frac{\partial}{\partial t}-\hat{H}(t)|\phi_0(t)\rangle,
\end{equation}
so the Schr\"{o}dinger equation is satisfied. By expressing $\hat{H}$ in terms of the coordinates $q_i$ in the above equation, we finally obtain
\begin{equation}
\theta_0(t)=-\int_0^t\frac{\dot{S}_2}{4(S_1^2-S_3^2)}dt=-\frac{1}{2}\tan^{-1}{(2S_2)},
\end{equation}
where in order to perform the integration we have exploited the constant of the motion (\ref{constant}).

At the boundary times $t_b=0, T$ we impose the frictionless conditions $[\hat{H}(t_b),\hat{I}(t_b)]=0$, so the system is driven from an eigenstate of $\hat{H}(0)$ to an eigenstate of $\hat{H}(T)$, which lead to the relations $\alpha(t_b)S_1(t_b)-JS_3(t_b)=0$, $\alpha(t_b)S_2(t_b)=0$, and $JS_2(t_b)=0$. Combining them with the initial conditions (\ref{Initial_Conditions}) we obtain $\alpha(0)=0$, $\alpha(T)/J=S_3(T)/S_1(T)$ and $S_2(T)=0$. Using additionally the value $\mathcal{N}$ of the desired final logarithmic negativity along with the expression (\ref{S3_negativity}) and the constant (\ref{constant}), we finally find
\begin{equation}
\label{Final_Conditions}
S_1(T)=\cosh{(\mathcal{N})}/2,\quad S_2(T)=0, \quad S_3(T)=\sinh{(\mathcal{N})}/2
\end{equation}
and
\begin{equation}
\label{boundary_alpha}
\alpha(0)=0,\quad \alpha(T)=J\tanh{(\mathcal{N})}.
\end{equation}
Having determined $\alpha(t)$ at the boundaries, we now move to find its intermediate values. If we take the time derivative of Eq. (\ref{system3}), then use Eq. (\ref{system2}) to eliminate $\dot{S}_2$, and finally express $S_1, S_2$ in terms of $S_3$ using Eqs. (\ref{constant}) and (\ref{system3}), we end up with the following differential equation for $S_3$
\begin{equation}
\label{S3_Dif_Equation}
\ddot{S}_3-4J\alpha(t)\sqrt{S_3^2+\dot{S}_3^2/(4J^2)+1/4}+4J^2S_3=0.
\end{equation}
The boundary conditions for $S_3$ can be found using Eqs. (\ref{Initial_Conditions}), (\ref{Final_Conditions}) and (\ref{boundary_alpha}). Since $S_2(0)=S_2(T)=0$, from Eq. (\ref{system3}) we obtain $\dot{S}_3(0)=\dot{S}_3(T)=0$. Since $\alpha(0)=0$ and $S_3(0)=0$, from Eq. (\ref{S3_Dif_Equation}) we find $\ddot{S}_3(0)=0$. Using the final values $S_3(T), \dot{S}_3(T), \alpha(T)$ in the same equation, we also find $\ddot{S}_3(T)=0$. The boundary conditions for $S_3$ are thus
\begin{equation}
\label{S3_initial}
S_3(0)=\dot{S}_3(0)=\ddot{S}_3(0)=0
\end{equation}
and
\begin{equation}
\label{S3_final}
S_3(T)=\sinh{(\mathcal{N})}/2,\quad \dot{S}_3(T)=\ddot{S}_3(T)=0.
\end{equation}
Following an inverse engineering approach \cite{Levy18}, we first pick a smooth function $S_3(t)$ satisfying the boundary conditions (\ref{S3_initial}), (\ref{S3_final}), and then find $\alpha(t)$ from Eq. (\ref{S3_Dif_Equation}). Choosing a polynomial ansatz for $S_3$ with six free coefficients, as many as the boundary conditions, we obtain
\begin{equation}
\label{interpolation}
S_3(s)=\frac{\sinh{(\mathcal{N})}}{2}(6s^5-15s^4+10s^3),\quad s=\frac{t}{T}.
\end{equation}
The input $\alpha(t)$ which accomplishes the desired shortcut is found from (\ref{S3_Dif_Equation}) to be
\begin{equation}
\label{alpha}
\alpha(t)=\frac{\ddot{S}_3+4J^2S_3}{4J\sqrt{S_3^2+\dot{S}_3^2/(4J^2)+1/4}}.
\end{equation}
Note that in theory, an arbitrarily large value of $\mathcal{N}$ can be obtained at any final time $T$. In practice, the larger is the desired final negativity, the larger is the maximum amplitude of the necessary control $\alpha(t)$. By setting an upper bound $\alpha(t)\leq A_0$, we limit the maximum achievable value of $\mathcal{N}$ for finite $T$.
On the other hand, if we require $\alpha(t)\geq 0$ throughout $0\leq t\leq T$, for implementation reasons, then this condition leads to a lower bound on the duration $T$, independent of the target logarithmic negativity $\mathcal{N}$. Indeed, using the polynomial form (\ref{interpolation}) of $S_3$ in the equivalent condition $\ddot{S}_3+4J^2S_3\geq 0$ for the numerator of (\ref{alpha}), the factors containing $\mathcal{N}$ can be omitted. Further manipulation of this inequality leads to the bound
\begin{equation}
\label{Minimum_Time}
T\geq(15/M)^{1/2}J^{-1}\approx 1.2527J^{-1},
\end{equation}
where
\begin{equation*}
M=\min_{1/2<s<1}\frac{s^2(6s^2-15s+10)}{(1-s)(2s-1)}\approx 9.558.
\end{equation*}

\begin{figure}[t]
\centering
\fbox{\includegraphics[width=0.9\linewidth]{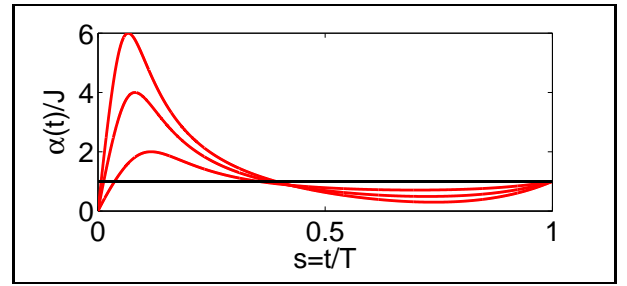}}
\caption{Bounded enhanced nonlinearity $\alpha(t)/J\leq A_0/J=2,4,6$ as a function of normalized time, corresponding to the three upper black circle points in Fig. \ref{fig:dissipation}.}
\label{fig:control}
\end{figure}

In Fig. \ref{fig:nodissipation} we plot as a function of the duration $T$ the maximum logarithmic negativity which can be achieved with the shortcut method when the control input is bounded from above, $\alpha(t)\leq A_0$. The three upper red lines correspond to different values of the maximum amplitude, $A_0/J=2,4,6$ from bottom to top, see Fig. \ref{fig:control} for the shape of $\alpha(t)$. The vertical black line in Fig. \ref{fig:nodissipation} indicates the critical duration (\ref{Minimum_Time}) beyond which it is also $\alpha(t)\geq 0$. Observe that a substantial improvement is obtained, compared to the case where $\alpha(t)$ is held constant (lower blue line), when the maximum amplitude $A_0$ is sufficiently larger than the coupling constant $J$. We can understand this behavior by using the simple differential equation (\ref{Diff_S2}) for $S_2$, which of course holds for constant $\alpha(t)$ but can be used to deduce some useful characteristics of the evolution for smooth enough $\alpha(t)$ like here. In the case of constant $\alpha(t)=\alpha_T<J$, observe from solution (\ref{solution_constant_a}) that $S_2$ is built and at the same time rotated towards $S_3$ in order to increase the logarithmic negativity (\ref{S3_negativity}), which is an increasing function of $|S_3|$. In the shortcut case with time-dependent $\alpha(t)$ displayed in Fig \ref{fig:control}, observe that there is a large enough time interval where $\alpha(t)>J$, above the horizontal black line. From Eq. (\ref{Diff_S2}) and the initial conditions for $S_2$ it turns that during this interval $S_2$ increases \emph{exponentially} in the negative direction. In the subsequent interval where $\alpha(t)<J$, this large (absolute) value of $S_2$ is rotated towards $S_3$, leading to an increased final negativity compared to the case with constant $\alpha(t)=\alpha_T<J$ throughout the process.


\textbf{Effect of dissipation}: We consider the simple dissipation model of Ref.\ \cite{Liew18}, where each second moment dissipates at a rate $\Gamma$. The system equations are modified as
\begin{subequations}
\label{system_dissipation}
\begin{eqnarray}
\dot{S}_1&=&-\Gamma S_1-2\alpha S_2+\Gamma/2,\label{d_system1}\\
\dot{S}_2&=&-2\alpha S_1-\Gamma S_2+2JS_3,\label{d_system2}\\
\dot{S}_3&=&-2JS_2-\Gamma S_3,\label{d_system3}
\end{eqnarray}
\end{subequations}
where note the differentiation of $S_1=\langle\hat{a}_1^\dagger\hat{a}_1+\hat{a}_2\hat{a}_2^\dag\rangle/2=\langle\hat{a}_1^\dagger\hat{a}_1+\hat{a}_2^\dag\hat{a}_2+1\rangle/2$.
Using the same shortcut inputs $\alpha(t)$ as in Fig. \ref{fig:nodissipation}, we simulate the above system for dissipation $\Gamma=0.1J$. The results are shown in Fig. \ref{fig:dissipation}. Observe that, in the presence of dissipation, the logarithmic negativity obtained with a bounded amplitude $\alpha(t)\leq A_0$ attains a maximum for a finite duration $T$ (black circles in the three upper red curves). For $A_0$ sufficiently larger than $J$, this maximum is larger than that obtained with constant $\alpha(t)=\alpha_T<J$ (black circle in the lower blue curve).

\textbf{Conclusion}: We have shown that using shortcuts to adiabaticity, the entanglement generated between two exciton-polariton cavities with a recently proposed method which effectively amplifies the system nonlinearity can be substantially enhanced. This work can find application in quantum information processing with polaritons, but also in other areas where nonlinear interacting bosons are encountered. %The results can be further optimized in the presence of dissipation \cite{Levy18}.

\noindent\textbf{Funding.} Greece and the European Union-European Regional Development Fund via the General Secretariat for Research and Technology (GRST) (project POLISIMULATOR).

%\section{References}

% Bibliography
%\bibliography{sample}

% Full bibliography added automatically for Optics Letters submissions; the following line will simply be ignored if submitting to other journals.
% Note that this extra page will not count against page length
%\bibliographyfullrefs{sample}

%Manual citation list

\end{document}